# Highly tunable hybrid metamaterials employing split-ring resonators strongly coupled to graphene surface plasmons

Peter Q. Liu,[1][†]* Isaac J. Luxmoore,[2][†]* Sergey A. Mikhailov,[3] Nadja A. Savostianova,[3] Federico Valmorra,[1] Jerome Faist,[1] Geoffrey R. Nash[2]

[1]Institute for Quantum Electronics, Department of Physics, ETH Zurich, Zurich CH-8093, Switzerland

[2]College of Engineering, Mathematics and Physical Sciences, University of Exeter, Exeter EX4 4QF, United Kingdom

[3]Institute of Physics, University of Augsburg, Augsburg 86159, Germany

[†]These authors contributed equally to the work.

*To whom correspondence should be addressed. E-mail: qliu@ethz.ch; i.j.luxmoore@exeter.ac.uk

**Abstract**

**Metamaterials and plasmonics are powerful tools for unconventional manipulation and harnessing of light. Metamaterials can be engineered to possess intriguing properties lacking in natural materials, such as negative refractive index. Plasmonics offers capabilities to confine light in subwavelength dimensions and to enhance light-matter interactions. Recently, graphene-based plasmonics has revealed emerging technological potential as it features large tunability, higher field-confinement and lower loss compared to metal-based plasmonics. Here, we introduce hybrid structures comprising graphene plasmonic resonators efficiently coupled to conventional split-ring resonators, thus demonstrating a type of highly tunable metamaterial, where the interaction between the two resonances reaches the strong-coupling regime. Such hybrid metamaterials are employed as high-speed THz modulators, exhibiting over 60% transmission modulation and operating speed in excess of 40 MHz. This device concept also provides a platform for exploring cavity-enhanced light-matter interactions and optical processes in graphene plasmonic structures for applications including sensing, photo-detection and nonlinear frequency generation.**

Since the turn of the century, research on metamaterials has progressed rapidly with substantial expansion of both the scope of novel functionalities and the operating frequency range enabled by different types of artificial structures[1-11]. Many of the demonstrated metamaterials are based on noble metals to take advantage of their negative permittivity below the plasma frequency. Real-time tunability of metamaterials is highly desired for many applications such as optical switches and modulators, however, it is a property lacking in metals. Different approaches have been developed to achieve tunable or reconfigurable metamaterials, among which several most effective realizations are based on changing the substrate properties[12-14]. Such approaches may find limitations where the properties of the substrate material cannot or should not be changed significantly, and metamaterial structures with intrinsic tunability[15-17] are keenly sought after. Graphene, a more recently discovered and intensively studied material with various interesting properties such as its tunable carrier density and high room-temperature carrier mobility, is a promising candidate for realizing tunable metamaterials across a broad spectral range[18]. Graphene's capability to support tightly confined surface plasmon (SP) in the terahertz (THz) to mid-infrared (MIR) spectral range[19,20] has been systematically investigated using scanning near-field optical microscopy[21-25] and demonstrated in various patterned graphene structures[26-33], including arrays of closely packed graphene ribbons (GR), which are essentially tunable metamaterials. However, the limited interaction between incident light and SP in monolayer graphene structures is not sufficient for many applications. Here, we demonstrate as proof-of-concept a type of electrostatically tunable hybrid metamaterial employing graphene plasmonic resonators strongly coupled to conventional metal-based metamaterials. In addition to their strong electromagnetic response and high tunability, such hybrid metamaterials also provide an interesting platform for exploring cavity-enhanced optical processes and light-matter interactions in graphene plasmonic structures for applications including sensing[34], photo-detection[35,36] and nonlinear frequency generation[37].

The proposed hybrid metamaterial concept can be applied to different types of structures, but in this work the specific realization is based on GRs and electric-field-coupled complementary split-ring resonators (C-SRRs)[38,39]. The schematics of a C-SRR unit cell, a GR and a unit cell of the proposed hybrid structure are illustrated in Fig. 1a-c. The rationale for such a hybrid structure design is the following: the near-field electric field (E-field) distribution associated with the LC-resonance of the C-SRR is highly localized and enhanced within the capacitor gap as shown by the simulation in Fig. 1d (see Methods), while the E-field distribution of the GR localized SP resonance is also highly confined in the vicinity of the GR (Fig. 1e). Both resonances are excited by E-field in the x-direction[26,38], and upon excitation their highly confined near-field also has the strongest E-field component in the x-direction. Therefore, embedding the GR in the middle of the C-SRR capacitor gap is an effective way to achieve strong near-field coupling of the two structures. When the localized SP resonance of the GR is tuned to approach and subsequently surpass the C-SRR LC-resonance by electrostatically varying the carrier density ($\omega_{SP} \propto n^{1/4} \propto |E_F|^{1/2}$), the spectral response of the hybrid metamaterial is modulated in the frequency range containing the two resonances. Such a mechanism of transmission modulation is significantly different from those employing tunable substrate free carrier absorption[12,13]. Due to the confined dimension along the GR width, the Drude-type free carrier absorption is considerably suppressed when the incident radiation is polarized perpendicular to the GR, and the localized SP resonance is the dominant process[26]. In order to achieve large modulation, the C-SRR capacitor gap should be designed to accommodate the GR with small margins to maximize the field overlap and hence the coupling strength. In addition, precise control of the hybrid metamaterial spectral response requires accurate information on the charge neutrality point (CNP) of the GRs. Therefore, a modified unit cell design is developed (Fig. 1f) in which a narrow gap along the horizontal symmetry axis is introduced to separate the C-SRR into two parts with minimal influence on its spectral response and field distribution (see Supplementary Information). This slight structural variation allows for convenient electrical characterization of the GR with the two C-SRR parts functioning as separate contacts, and may also enable direct probing of the GR photo-response[35,36] as influenced by the C-SRR cavity.

Following the above design principle, we have developed and experimentally investigated multiple different hybrid metamaterial structures targeting two different operating frequency ranges, i.e. ~10 THz and ~4.5 THz. To achieve this, GR widths of ~400 nm and ~1.8 μm, respectively, are chosen to ensure that the SP resonance can be electrostatically tuned across the C-SRR LC resonance[31], even when moderate screening effect from the metal is present (see Supplementary Information). The GRs are implemented with large-area monolayer graphene grown by chemical vapor deposition (CVD) and transferred onto a $SiO_2$/Si substrate, which is also utilized as the back-gate for electrostatic control of the graphene carrier density. Electron-beam lithography is employed for patterning the structures (see Methods). Figure 2 shows scanning electron microscope (SEM) images of a fabricated C-SRR-GR array designed to operate around 10 THz.

To investigate the carrier density dependent spectral response of the fabricated C-SRR-GR hybrid metamaterial devices, their transmission spectra are characterized employing Fourier transform spectroscopy (FTIR) with the normally incident radiation polarized perpendicular to the GRs (see

Methods). Figure 3 summarizes the key results from two hybrid metamaterial devices (HM1 and HM2) designed to operate around 10 THz. The two devices employ GRs of identical width (400 nm), while the C-SRR unit cell of HM2 is scaled up in-plane by 10% compared to that of HM1 to attain a slightly lower (by ~10%) LC-resonance frequency. Figure 3a shows the transmission spectra of HM1 at 3 different carrier densities in comparison with that of a reference bare C-SRR array. The bare C-SRR array exhibits two resonances in the measured frequency range. The resonance at ~10 THz corresponds to the designed LC-resonance, whereas the resonance at ~16 THz is attributed to the complementary of the half-wave dipole antenna associated with the vertical dimension of the original SRR[12]. Figure 3a also shows the transmission spectrum of an array of 400-nm-wide GRs with a carrier density of $2.3 \times 10^{13}$ cm$^{-2}$, in which two resonances originating from the strong coupling of the intrinsic graphene SP resonance to an optical phonon mode of the underlying SiO$_2$ layer are present[31] (see Supplementary Information). Both the C-SRR LC-resonance and the GR SP resonance at ~10 THz have a quality factor larger than 3, the latter of which also benefits from the coupling to the SiO$_2$ phonon. Although the higher-frequency graphene SP resonance interacts with the higher-frequency C-SRR resonance, their coupling strength is expected to be weak due to the large spatial extent of the half-wave dipole antenna mode, and therefore is not the focus of this study. Focusing on the spectral range near the LC-resonance, it is evident that at the CNP, the transmission spectrum of the device is close to that of the corresponding reference C-SRR array, whereas large transmission modulation in both amplitude and line shape is realized by varying the carrier density, with up to ~50% relative modulation of the peak transmission achieved with HM1. This is in sharp contrast to previous investigations where a continuous layer of graphene is brought into contact with SRRs[40,41] and the LC-resonance of the SRRs is considerably diminished even at the CNP condition[41].

Moreover, the transmission spectra at high carrier densities display a double-peak feature. This is a characteristic feature of the strong coupling[42,43] between the C-SRR LC-resonance and the GR localized SP resonance due to efficient near-field interaction, which leads to two hybridized modes, i.e. polaritons. The evolution of the transmission of HM1 with increasing carrier density is more clearly revealed in Fig. 3b, in which the higher-frequency peak is observed to gradually blue-shift with decreasing strength, while the lower-frequency peak emerges and becomes more pronounced at higher carrier densities. Similar behavior is also observed in HM2 (Fig. 3c) with a noteworthy difference that the lower-frequency peak becomes more dominant at the highest achieved carrier densities. This is consistent with the fact that the LC-resonance frequency is lower in HM2 which allows the SP resonance to surpass the LC-resonance by a larger amount. The frequencies of the individual peaks in the transmission spectra are extracted by fitting the data with two Lorentzian functions, and the results are plotted in Fig. 3d for HM1 and Fig. 3e for HM2. The evolution of the upper and the lower polariton branches with the SP resonance frequency are well fitted with a formula

$$\omega_\pm^2 = \frac{\omega_{C-SRR}^2 + \omega_{SP}^2}{2} \pm \sqrt{\left(\frac{\omega_{C-SRR}^2 - \omega_{SP}^2}{2}\right)^2 + \beta^2 \omega_{C-SRR}^2 \omega_{SP}^2} \qquad (1)$$

describing the anti-crossing behavior of two strongly coupled resonances[44], where $\beta$ is a parameter determining the coupling strength, and may also be interpreted with an equivalent circuit model as the relative strength of a capacitive coupling between two LC-resonators (see Supplementary Information). The splitting between the two branches ( $\omega_+ - \omega_-$ ) when $\omega_{SP} = \omega_{C-SRR}$ is approximately $\beta \omega_{C-SRR}$, and the coupling strength (half of this splitting) is found to be ~1.1 THz in both devices, more than 10% of the individual resonance frequency and is therefore also within the ultra-strong coupling regime[45].

These experimental observations are well reproduced by full-wave simulations. Figure 4a shows the simulated transmission spectra of HM1 at various carrier densities in comparison with that of a bare C-SRR array. Both the large transmission modulation and the double-peak profile at high carrier densities are consistent with the experimental observations. The simulated E-field profiles associated with the two transmission peaks at $|E_F|$=0.4 eV (Fig. 4b-c) evidently reveal that the lower-frequency peak stems from a bonding mode, and the higher-frequency peak an anti-bonding mode[46]. The two hybridized modes exhibit a typical anti-crossing behavior (Fig. 4d) as well as exchange of their oscillator strengths, as the localized SP resonance approaches and subsequently surpasses the LC-resonance with increasing carrier density. The coupling strength is extracted to be ~1.15 THz (see Supplementary Information), in excellent agreement with the experimental result (~1.1 THz). The field localization and enhancement near the GR in the hybrid structure is also considerably increased compared to either the bare C-SRR or the bare GR (the field enhancement near the GR edges is ~8 times higher when the C-SRR is present), which leads to enhanced absorption of incident radiation and may find applications in chemical and biological sensing. Furthermore, with graphene of higher material quality and thus higher carrier mobility, the carrier density dependent modulation of the transmission is expected to be further improved. As shown in the simulated spectra in Fig. 4e assuming a carrier mobility that is realistic for the state-of-the-art CVD grown graphene[47,48], the transmission of such a hybrid metamaterial can be switched almost completely off across a wide frequency range by controlling the graphene carrier density. Such a superior modulation performance is a direct consequence of the strong coupling between the two resonances.

The demonstrated C-SRR-GR hybrid metamaterials can be utilized as efficient modulators and switches. Since fast modulation is highly desired for many applications such as real-time compressive imaging[11,49] and wireless communication, the modulation speed of several devices operating around 4.5 THz are investigated using a THz quantum cascade laser (4.7 THz) as the source. Figure 5a-b show the transmission spectra of two hybrid metamaterial structures (HM3 and HM4) operating at ~4.0 THz and ~4.8 THz, respectively, at various back-gate voltages in comparison with the transmission spectrum of the corresponding reference bare C-SRR array. The transmission exhibits similar (even higher) carrier density dependent modulation as observed in the 10 THz devices, with over 60% relative modulation of the peak transmission. The plateau-shaped transmission spectra in both figures correspond to the situation where the two hybridized modes have similar strengths, and they also suggest that the two resonances are in the critical coupling regime, i.e., the coupling strength is close to half of the broadening, and thus the individual transmission peaks associated with each polariton mode are not as clearly resolved as in the 10 THz devices, consistent with the simulation (see Supplementary Information). This is a result of the relatively lower quality factor of both the GR SP resonance and the C-SRR LC-resonance in this frequency range (Q~2). The C-SRR-GR structure allows for straightforward parallel electrical connection of all the unit cells, as illustrated in Fig. 5c, to minimize the total resistance of the device. Hence, as the total capacitance (C) increases proportional to the device area, the total resistance (R) associated with the entire C-SRR-GR array scales inversely proportional to the device area, facilitating high-speed operation of large-area devices. Figure 5d shows the modulation speed measurement (see Methods) of a HM3 device (1 mm × 1 mm area) and a HM4 device (0.5 mm × 0.5 mm area), respectively. The measured 3 dB cut-off frequency is ~19 MHz for the larger device and ~41 MHz for the smaller device. To the best of our knowledge, such performance is superior to the state-of-the-art for fast tunable metamaterials in the literature which was achieved with much smaller device area[50], and is several times higher than that reported for devices with similar area[49]. Electrical characterizations show that the resistance of the C-SRR-GR array is indeed not the limiting factor for the RC constant of these devices, whereas the dominant resistance contribution is from the low-doped Si substrate (due to the in-plane current flow) and the input resistance (50 Ω) of the driving voltage source (see Supplementary Information). The modulation speed can be further enhanced up to GHz range without the need of reducing the device area. For example, by utilizing a wire-grid contact on the backside of the Si substrate, the substrate contribution to the total resistance can be reduced by more than one order of magnitude without affecting the transmission of THz radiation. The total capacitance can also be significantly reduced with further optimization of the device architecture, such as employing a transparent local top-gate for only the GRs or using resonator structures covering less area.

The demonstrated concept of coupling graphene-based plasmonic structures with conventional metal-based metamaterials to achieve highly tunable hybrid metamaterials can be straightforwardly extended to other frequency ranges (e.g. MIR) and different graphene plasmonic and/or metameterial structures (e.g. graphene disks and negative-index metamaterials), to further broaden the scope of novel functionalities. The presented device structures also provide a platform for further exploring intriguing cavity-enhanced light-matter interactions and optical processes in graphene plasmonic structures which may lead to various additional applications such as sensing, photo-detection and non-linear frequency generation.

**Methods**

**Device fabrication.** The C-SRR-GR hybrid metamaterials are fabricated from pre-transferred large-area CVD grown monolayer graphene on 300 nm thick $SiO_2$, which has an underlying Si substrate with resistivity of ~10 Ωcm. Electron-beam lithography and reactive ion etching with $O_2$ plasma are used to pattern the continuous graphene sheet into arrays of ribbons with the designed widths and separations. A second stage of electron-beam lithography followed by deposition of Cr/Au (thickness 5/80 nm) and a lift-off process defines the C-SRR arrays, which are in direct contact with the GRs to enable electrostatic control of the carrier density using the Si substrate as the back-gate, as well as direct characterization of the electrical properties of the GRs. The devices operating around 10 THz have a surface dimension of 0.6 mm×0.6 mm, and the devices operating around 4.5 THz have various surface dimensions: 2 mm×2 mm, 1 mm×1 mm or 0.5 mm×0.5 mm.

**Transmission characterization.** Transmission spectra of the C-SRR-GR hybrid metamaterial devices operating at two different frequency ranges are also characterized with two different experimental setups. Transmission characterization of the 10 THz devices are performed at room temperature in the ambient condition with a FTIR coupled to an infrared microscope. The normally incident radiation is linearly polarized perpendicular to the GRs using a broadband wire grid polarizer which limits the low-frequency measurement range to ~6.7 THz. The transmission spectra of the ~4.5 THz devices are characterized at room temperature in the vacuum chamber of another FTIR. The normally incident radiation is also linearly polarized perpendicular to the GRs using another broadband wire grid polarizer with broader operating frequency range. The low-frequency measurement range is limited to ~1 THz by the detector.

**Modulation speed measurement.** The modulation speed measurement is conducted with a 4.7 THz quantum cascade laser as the light source, a function generator as the voltage source for modulating the hybrid metamaterial devices (up to 80 MHz), and a superconducting hot electron bolometer as the fast detector (with response up to 200 MHz), feeding the output signal to a lock-in amplifier with demodulation frequency up to 50 MHz (limiting the frequency range of the measurement). The output voltage from the function generator is set to be a sinusoidal signal with 10 V amplitude (limited by the equipment). The modulation depth at any specific frequency in Fig. 5c is normalized to the value measured at 1 MHz modulation frequency.

**Full-wave simulation.** The spectral responses and field distributions of all the investigated structures are simulated using finite-element frequency domain methods with CST Microwave Studio. The mono-layer graphene sheet is modeled as a 0.3 nm thin layer with a dynamical surface conductivity described by the Drude model as $\sigma(\omega) = \frac{e^2 E_F}{\pi \hbar^2} \frac{i}{\omega + i\tau^{-1}}$, which is an accurate approximation when the Fermi energy $E_F$ is significantly higher than the corresponding energy of the frequency range investigated. In most of the simulations, the carrier relaxation time is assumed to be 50 fs which is realistic for the graphene material available for our experimental demonstration[31]. The frequency dependent permittivity of the $SiO_2$ layer (300 nm thick) is computed taking into account the surface optical phonon mode at ~14.5 THz (see Supplementary Information), while tabulated data from ref 51 are used for the low-doped Si substrate (modeled as 5 μm thick). The C-SRR structure is modeled to consist of 100 nm thick gold, with the permittivity described by the Drude model assuming the plasma frequency of $2\pi \times 2.184 \times 10^{15}$ s$^{-1}$ and the damping constant of $2\pi \times 1.7 \times 10^{13}$ s$^{-1}$.

**Acknowledgement**

We would like to thank K. Otani for providing the THz quantum cascade laser used in the measurement of the device modulation speed, C. Maissen and G. Scalari for useful discussions and comments on the manuscript. This work was supported by the European Union under the FET-open grant GOSFEL, the ERC Advanced Grant MUSiC, the Graphene Flagship (No. CNECT-ICT-604391), and the Swiss National Science Foundation through NCCR QSIT. G.R.N. also acknowledges the support of the UK Engineering and Physical Sciences Research Council through a Fellowship in Frontier Manufacturing.

**Author contributions**

P.Q.L. and I.J.L. conceived the experiment. P.Q.L. designed the device structure and performed the simulation. I.J.L. fabricated the devices. P.Q.L. and I.J.L. conducted the measurements. All authors contributed to the data analysis and discussions. S.A.M. and N.A.S. performed theoretical calculations. P.Q.L. and J.F. developed the equivalent circuit model. P.Q.L., I.J.L., S.A.M. and N.A.S. wrote the manuscript, with input from all authors.

**Competing financial interests**

The authors declare no competing financial interests.

**Figure legends**

**Figure 1 | Design of the tunable C-SRR-GR hybrid metamaterials. a,** Schematic of a typical C-SRR. **b,** Schematic of a GR on a substrate. **c,** Schematic of a typical unit cell of the proposed C-SRR-GR hybrid metamaterials. **d,** Simulated x-component of the E-field distribution associated with the LC-resonance of the C-SRR in (**a**). **e,** Simulated x-component of the E-field distribution associated with the localized SP resonance of the GR in (**b**). **f,** Schematic of a typical unit cell of the proposed C-SRR-GR hybrid metamaterials employing a two-piece C-SRR design.

**Figure 2 | SEM images of a fabricated C-SRR-GR hybrid metamaterial device (HM1). a,** SEM image of an array of the C-SRR-GR unit cells. **b,** Close-up SEM image of a single C-SRR-GR unit cell.

**Figure 3 | Transmission modulation of C-SRR-GR hybrid metamaterials operating at ~10 THz. a,** Transmission spectra (normalized to the substrate transmission) of the C-SRR-GR hybrid metamaterial HM1 at different graphene carrier densities in comparison with the transmission spectrum of a reference bare C-SRR array. The transmission spectrum of a GR array exhibiting the localized SP resonances is also shown. **b,c,** Stacked transmission spectra (5% offset) at varied carrier densities (from the CNP to ~$2.3\times10^{13}$ cm$^{-2}$) of two C-SRR-GR hybrid metamaterial devices, HM1 (**b**) and HM2 (**c**), which employ slightly different C-SRR designs. The colored curves in (**b**) correspond to those in (**a**) of the same color. **d,e,** Symbols are the extracted peak frequencies of the two hybridized modes of devices HM1 (**d**) and HM2 (**e**) as a function of the carrier density ($n^{1/4}$) from the corresponding transmission spectra in (**b**) and (**c**), respectively. The solid curves are fits of the data points with equation (1) describing the anti-crossing of two strongly coupled resonances, i.e. the localized SP resonances in the GRs (blue dashed line) and the C-SRR LC-resonance (green dashed line). See Supplementary Information for details of the fitting procedure.

**Figure 4 | Simulated spectral response of the tunable C-SRR-GR hybrid metamaterial operating at ~10 THz. a,** Simulated transmission spectra of the C-SRR-GR hybrid metamaterial HM1 at various graphene Fermi energies (carrier densities) in comparison with that of a reference bare C-SRR array. The carrier relaxation time of 50 fs is used in this simulation. **b,c,** Simulated z-component of the E-field distributions associated with the two hybridized modes (transmission peaks), i.e. the bonding mode (**b**) and the anti-bonding mode (**c**), at $|E_F|$=0.4 eV. The upper graphs correspond to the field distributions in the x-y-plane right below the C-SRR-GR unit cell, and the lower graphs correspond to those in the x-z-plane indicated by the dashed line in the upper graphs. **d,**

Symbols are the extracted peak frequencies of the two hybridized modes in (**a**) as a function of the graphene carrier density ($n^{1/4}$). The solid black curves are fits of the data points using equation (1), with the dashed lines of the same meaning as in Fig. 3d-e. The fits of the corresponding experimental data shown in Fig. 3d are also plotted in solid grey curves for comparison. **e**, Simulated carrier-density-dependent transmission spectra of a C-SRR-GR hybrid metamaterial in comparison with that of a reference bare C-SRR array, assuming 1 ps for the graphene carrier relaxation time.

**Figure 5 | Transmission modulation of C-SRR-GR hybrid metamaterials operating around 4.5 THz. a,b,** Transmission spectra (normalized to the substrate transmission) of the C-SRR-GR hybrid metamaterial devices HM3 (**a**) and HM4 (**b**) at different back-gate voltages (graphene carrier densities) in comparison with the transmission spectrum of the corresponding reference bare C-SRR array. The inset in each graph shows a SEM image of a device unit cell. The vertical gray dashed lines indicate the frequency of the THz quantum cascade laser used in the modulation speed measurement. **c,** Schematic of the parallel electrical contact for all the C-SRR-GR unit cells. **d,** Symbols are the normalized modulation depth of two hybrid metamaterial devices, HM3 (top) and HM4 (bottom), as a function of modulation frequency. Solid curves are the fits of the measured data based on a standard RC circuit model.

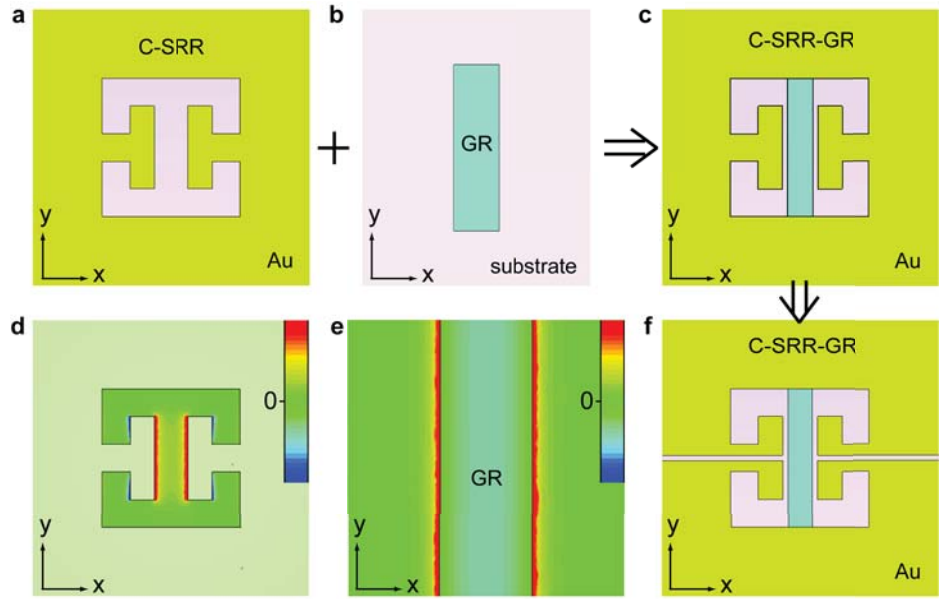

Figure 1

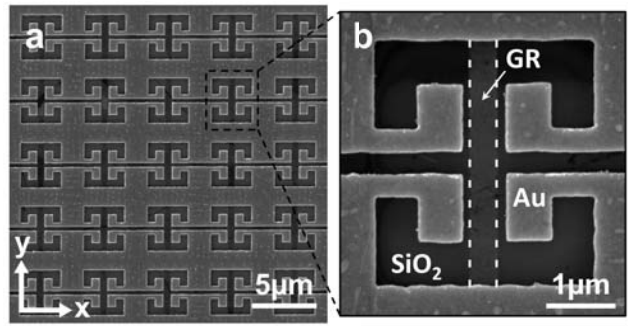

Figure 2

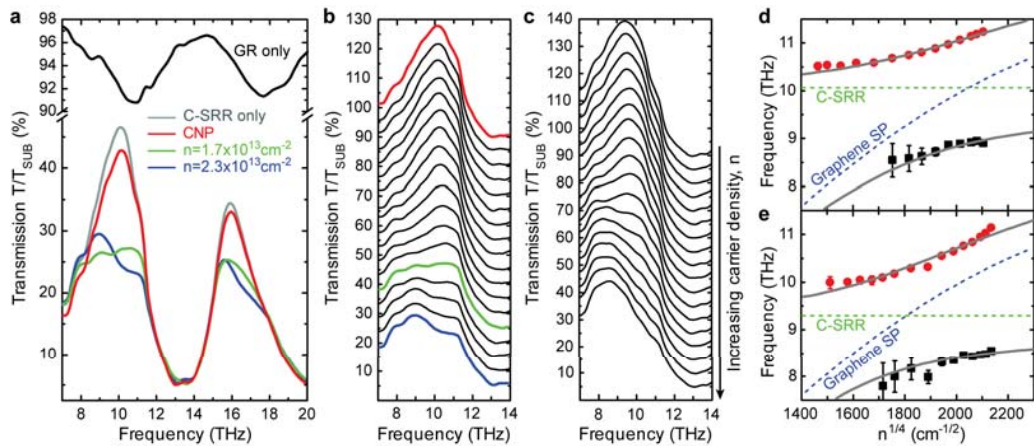

Figure 3

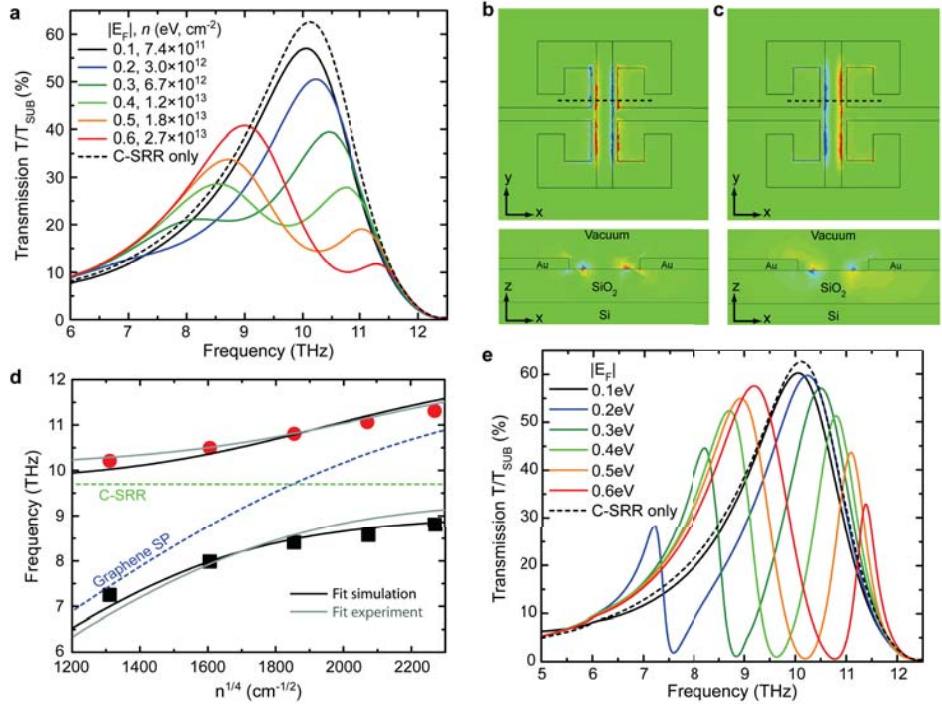

Figure 4

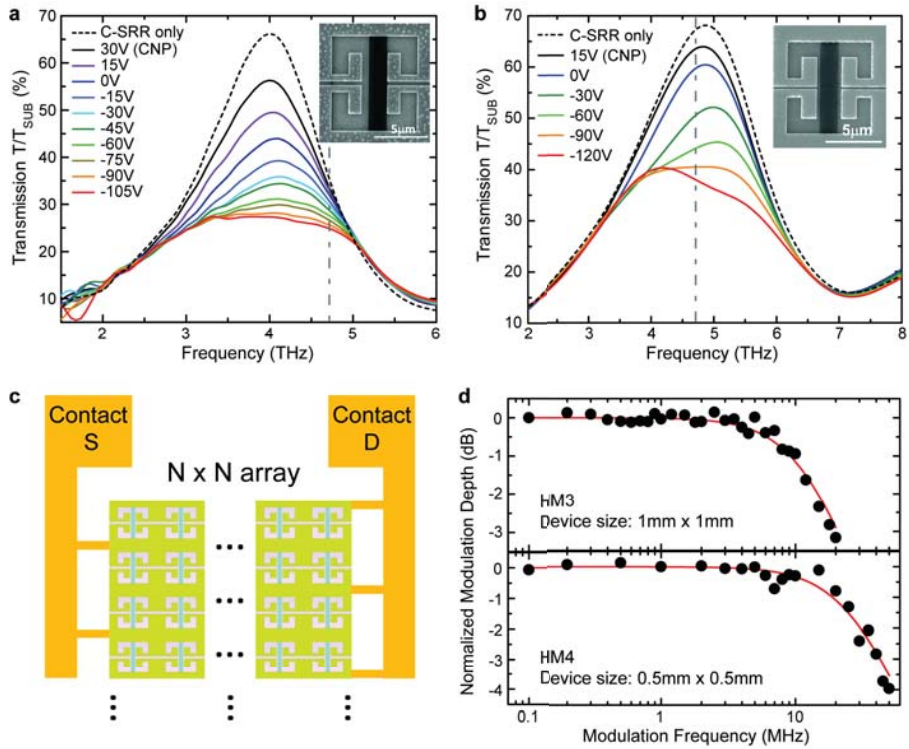

Figure 5

# Highly tunable hybrid metamaterials employing split-ring resonators strongly coupled to graphene surface plasmons

Peter Q. Liu,[1][†]* Isaac J. Luxmoore,[2][†]* Sergey A. Mikhailov,[3] Nadja A. Savostianova,[3] Federico Valmorra,[1] Jerome Faist,[1] Geoffrey R. Nash[2]

[1]Institute for Quantum Electronics, Department of Physics, ETH Zurich, Zurich CH-8093, Switzerland
[2]College of Engineering, Mathematics and Physical Sciences, University of Exeter, Exeter EX4 4QF, United Kingdom
[3]Institute of Physics, University of Augsburg, Augsburg 86159, Germany
[†]These authors contributed equally to the work.
*To whom correspondence should be addressed. E-mail: qliu@ethz.ch; i.j.luxmoore@exeter.ac.uk

## S1. Comparison between a conventional C-SRR and a modified two-piece C-SRR designs.

By introducing a narrow gap along the horizontal symmetry axis of a conventional C-SRR unit cell, the C-SRR is electrically separated into two parts with minimal influence on the near-field E-field distribution (as indicated by the comparison between Fig. S1a and Fig. S1b) and the transmission spectrum (shown in Fig. S1c).

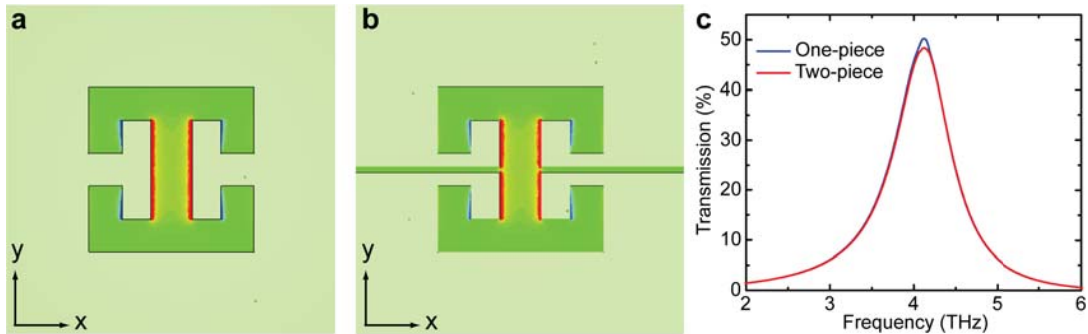

**Figure S1 | Simulated E-field distributions and transmission spectra of two different C-SRR structures. a,** Simulated x-component of the near-field E-field distribution of a conventional (one-piece) C-SRR design. **b,** Simulated x-component of the near-field E-field distribution of a two-piece C-SRR design. **c,** Comparison of the transmission spectra of the two C-SRR designs in (**a**) and (**b**).

## S2. Screening of the GR localized plasmons by nearby metallic structures.

If the distance between the edges of a GR and the nearby metallic structures (such as the C-SRR in this work) is sufficiently small compared to the thickness of the metallic structures, the localized SP resonance frequency of the GR is reduced as a result of the screening effect from the metal. The collective oscillation of the charge carriers (i.e. the SP resonance) in a GR is a consequence of the restoring force

$$F_x(x) \propto \frac{eQ_l}{\epsilon}\left(\frac{1}{|x + W/2|} + \frac{1}{|x - W/2|}\right), \quad |x| < \frac{W}{2},$$

(S1)

arising due to the charge density fluctuations with the linear charge density $Q_l$, localized near the GR edges and indicated by the positive (+) and negative (-) signs in Fig. S2. In the presence of nearby thick (compared to the distances $d_1$ and $d_2$) metallic structures which may be approximated as infinitely high metallic walls extending in both directions, this restoring force is modified by the image charges induced in the metal,

$$F_x(x) \propto \frac{eQ_l}{\epsilon}\left(\frac{1}{|x+W/2|} + \frac{1}{|x-W/2|} - \frac{1}{|x+W/2+2d_1|} - \frac{1}{|x-W/2-2d_2|}\right), \quad |x| < \frac{W}{2}.$$
(S2)

As a result, the SP resonance frequency is reduced to

$$\omega'_{SP} = \omega^0_{SP}\sqrt{1 - \frac{1}{2(1+\alpha_1)} - \frac{1}{2(1+\alpha_2)}},$$
(S3)

where $\omega^0_{SP}$ is the unscreened SP resonance frequency of the GR and $\alpha_{1,2} = 4d_{1,2}/W$. This calculation provides the theoretical upper limit of the screening effect induced by nearby metallic structures, while the actual reduction of the SP resonance frequency depends on the geometrical details of the structures and can be computed numerically.

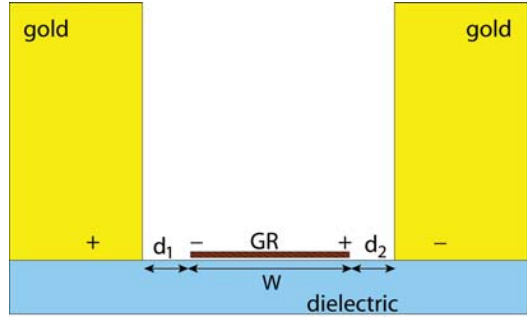

**Figure S2 | Schematic of a GR placed between two relatively thick metallic (gold) structures with narrow gaps.** Induced image charges are represented by the positive (+) and negative (-) signs in the metallic structures.

## S3. Fitting procedure of the hybridized modes of the C-SRR-GR hybrid metamaterials.

The efficient coupling of the C-SRR LC resonance to the GR localized SP resonance leads to the observed hybridized modes in the transmission spectra of the C-SRR-GR hybrid metamaterials. To quantitatively analyse the carrier density dependent spectral response of these hybrid metamaterials, we have also investigated the tuning of the SP resonances in bare GR arrays (without the C-SRRs). The GR arrays are fabricated on the same $SiO_2$/Si substrates used for the hybrid metamaterials, which allow for tuning of the carrier density via the applied back-gate voltage [1]. $SiO_2$ is a polar dielectric and as such, supports surface optical (SO) phonons, which can couple to SP resonances in the GRs via the long-range Fröhlich interaction [2,3], resulting in multiple surface-plasmon-phonon-polariton (SP3) modes [1,4]. The transmission spectrum of such an array of 400-nm-wide GRs is measured as a function of carrier density and plotted in Fig. S3a. In the measured frequency range, two resonances are observed, labelled $P_1$ and $P_2$, which originate from the interaction of the graphene SP and the SO phonon at ~14.5 THz [1]. As the carrier density is increased, there is a monotonic increase in the frequency of both resonances, with the peak positions plotted as a function of carrier density ($n^{1/4}$) in Fig. S3b.

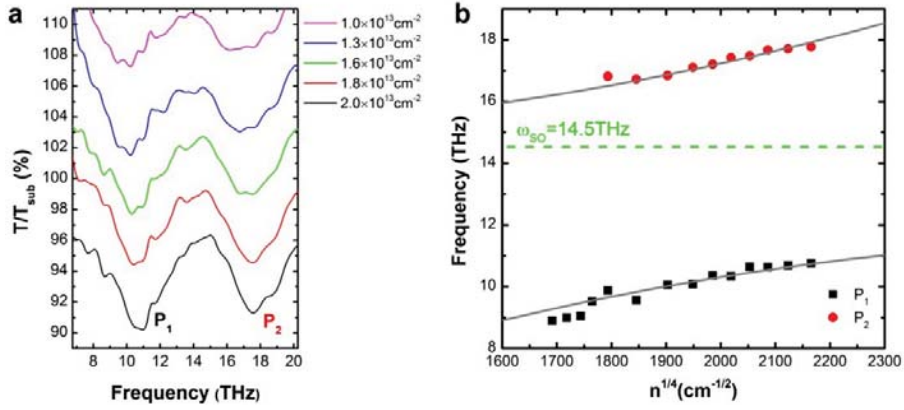

**Figure S3 | Experimental investigation of a reference array of bare GRs. a,** Transmission spectra of an array of GRs (400 nm wide) normalised to the substrate transmission for different values of the carrier density. The labels $P_1$ and $P_2$ indicate the two SP3 modes. The curves are offset by 2%. **b,** Extracted peak frequencies for $P_1$ and $P_2$ plotted as a function of $n^{1/4}$. The solid grey lines show a fit to the experimental data with Eq. S12.

To derive an equation to describe the anti-crossing of the plasmon and SO phonon modes, we consider a 2D layer with the conductivity $\sigma(\omega)$ lying on the boundary $z = 0$ of two media with the bulk dielectric functions $\epsilon_1(\omega)$ and $\epsilon_2(\omega)$. The spectrum of electromagnetic waves running along this boundary in the $x$-direction and localized at this boundary is determined by the dispersion equation (in Gaussian units)

$$\varepsilon_1(\omega) + \varepsilon_2(\omega) + \frac{4\pi i |q| \sigma(\omega)}{\omega} = 0 .$$

(S4)

The conductivity can be described using the Drude Model, which for the case of graphene and neglecting the carrier relaxation is

$$\sigma(\omega) = \frac{e^2 v_F}{\hbar} \sqrt{\frac{n}{\pi}} \frac{i}{\omega},$$

(S5)

where $n$ is the sheet carrier density and $v_F$ is the Fermi velocity. We assume the first medium is vacuum, $\epsilon_1(\omega) = 1$ and the second has the dielectric function $\epsilon_2(\omega)$ and consider several special cases. Firstly, the second medium is also vacuum $\epsilon_2(\omega) = 1$. The dispersion equation then gives

$$\omega^2 = \frac{2\pi |q| e^2 v_F}{\hbar} \sqrt{\frac{n}{\pi}} \equiv \omega_p^2 ,$$

(S6)

where we define $\omega_p$ as the plasma frequency for graphene in vacuum. In the second case, we assume there is no graphene, i.e. $n = 0$, and the second medium is a polar semiconductor with

$$\epsilon_2(\omega) = \epsilon_\infty - (\epsilon_0 - \epsilon_\infty) \frac{\omega_{TO}^2}{\omega^2 - \omega_{TO}^2},$$

(S7)

where $\omega_{TO}$ is the frequency of the transverse optical phonons, and $\epsilon_0$ and $\epsilon_\infty$ are the dielectric constants at frequencies much lower and much higher than $\omega_{TO}$, respectively. The dispersion equation gives

$$1 + \epsilon_2(\omega) = 1 + \epsilon_\infty - (\epsilon_0 - \epsilon_\infty)\frac{\omega_{TO}^2}{\omega^2 - \omega_{TO}^2} = 0,$$

(S8)

and can be rewritten as

$$\omega^2 = \omega_{SO}^2 = \omega_{TO}^2 \frac{\epsilon_0 + 1}{\epsilon_\infty + 1},$$

(S9)

where $\omega_{SO}^2$ is the frequency of the SO phonons. Therefore, the dispersion equation at the interface of a vacuum and a polar semiconductor can be written as

$$1 + \epsilon_2(\omega) = (1 + \epsilon_\infty)\frac{\omega^2 - \omega_{SO}^2}{\omega^2 - \omega_{TO}^2} = 0.$$

(S10)

Finally, for the case of graphene on the surface of a polar crystal, the dispersion equation is

$$1 + \epsilon_2(\omega) - 2\frac{\omega_p^2}{\omega^2} = (1 + \epsilon_\infty)\frac{\omega^2 - \omega_{SO}^2}{\omega^2 - \omega_{TO}^2} - 2\frac{\omega_p^2}{\omega^2} = 0,$$

(S11)

which leads to the following biquadratic equation, where the two solutions describe the coupled plasmon-phonon modes,

$$\omega_{SP,\pm}^2 = \frac{\omega_{SO}^2 + \frac{2\omega_p^2}{\epsilon_\infty + 1}}{2} \pm \frac{\sqrt{\left(\omega_{SO}^2 - \frac{2\omega_p^2}{\epsilon_\infty + 1}\right)^2 + 8\omega_{SO}^2\omega_p^2\left(\frac{1}{\epsilon_\infty + 1} - \frac{1}{\epsilon_0 + 1}\right)}}{2}.$$

(S12)

To fit the experimental data shown in Fig. S3b, we use Eq. S12 with $\omega_p$ defined according to Eq. S6 with a slight modification as

$$\omega_p = \sqrt{\frac{2\pi|q|e^2 v_F}{\hbar\sqrt{\pi}}} n^{1/4} \to A\sqrt{\frac{2\pi^2 e^2 v_F}{\sqrt{\pi}\hbar W}} n^{1/4},$$

(S13)

where $A$ is a fitting parameter to take into account the relation between the frequency of the localized SP resonance in a GR and the frequency of a SP wave with the wave vector |q| in a continuous graphene sheet (Eq. S6); the factor $A$ was numerically found to be $A\sim 0.86$ in [5], see Figure 1 there (see also [6]). The solid lines in Fig. S3b show a fit to the experimental data using Eq. S12, with $\omega_{SO}$ = 14.5 THz (ref 1), $\epsilon_0$=3.9 and the fit parameters $\epsilon_\infty$=2.81 and A=0.77. The lower branch, $\omega_{SP,-}$ is subsequently used to fit the anti-crossing of the C-SRR-GR devices, HM1 and HM2, employing Eq. 1 in the main text with $\omega_{SP} = \omega_{SP,-}$, and with $\omega_{SO} = 14.5$ THz, $\epsilon_0=3.9$, $\epsilon_\infty=2.81$ as fixed parameters and A, β, $\omega_{C-SRR}$ as fitting parameters. The same fitting procedure is carried out for the simulated transmission spectra of HM1 in Fig. 4 of the main text. Table S1 lists the fit parameters for the two devices, HM1 and HM2, in comparison with those for the simulated spectra of HM1. Excellent agreement is found between the experimental results and the simulation.

Table S1 | C-SRR-GR fit parameters.

|  | HM1 | HM2 | Simulation (HM1) |
|---|---|---|---|
| A | 0.72 | 0.72 | 0.75 |
| β | 0.22 | 0.23 | 0.24 |
| $\omega_{C-SRR}$ (THz) | 10.1 | 9.3 | 9.7 |

It is worth pointing out that when evaluating the experimental carrier density of the 400 nm wide GRs at a given back-gate voltage, the parallel-plate capacitor model is not a good approximation in this case, as the GR width is comparable to the thickness $d$ (~300 nm) of the dielectric layer and the edge effect is hence significant. To this end we consider a thin conducting layer of the width $W$ (the GR) lying on a dielectric slab at a distance $d$ from the grounded conducting substrate. The distribution of the electric potential $\phi$ in the free space and in the dielectric is described by the Laplace equation with the boundary conditions $\phi = 0$ at the grounded substrate and $\phi = V_G$ at the GR. Solving this equation with the Green function technique, we reduce the problem to an integral equation for the electron density $n(x)$ inside the GR

$$V_G = \int_{-\infty}^{\infty} \left( \frac{2e}{q(1 + \epsilon \coth qd)} \int_{-W/2}^{W/2} n(x')\cos(q_x x')dx' \right) \cos(q_x x) \, dq_x, \quad |x| < \frac{W}{2}, \tag{S14}$$

where $q = |q_x|$. Expanding the density $n(x)$ in a series of Chebyshev polynomials $T_n(2x/W)$ we reduce the problem to an infinite set of algebraic equations for the expansion coefficients. Truncating this set (the convergence is checked and found to be very fast) we get the relation between the average density of electrons in the GR $n = \langle n(x) \rangle$ and the gate voltage. The resulting ratio of the capacitances $C_W/C_\infty$ as a function of $d/W$ at $\epsilon = 3.9$ (SiO$_2$) is shown in Fig. S4, where $C_\infty$ is the capacitance predicted by the parallel-plate capacitor model.

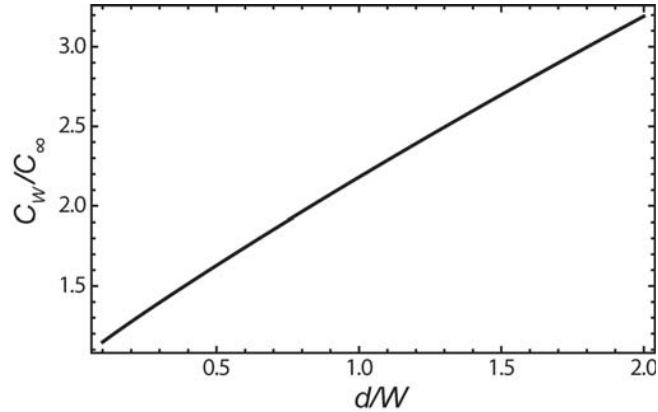

**Figure S4 | The calculated ratio $C_W/C_\infty$ as a function of $d/W$ at $\epsilon = 3.9$.**

## S4. Equivalent circuit model of the C-SRR-GR hybrid metamaterials.

We introduce an equivalent circuit model as an alternative description of the operating mechanism of the C-SRR-GR hybrid metamaterials, which also provides a more intuitive interpretation of the coupling strength characterized by the parameter $\beta$ in Eq. 1 in the main text. The equivalent circuit diagram for the C-SRR-GR hybrid structure is illustrated in Fig. S5 (enclosed by the dashed square), in which both the C-SRR and the GR are modeled as LC-circuits (losses are neglected for simplicity). The GR is modeled as a parallel LC-circuit because the SP resonance corresponds to enhanced reflection, while the C-SRR is modeled as a series LC-circuit due to enhanced transmission at the resonance. Since the coupling between the C-SRR and the GR is established via the interaction between the near-field electric field of one structure and the charge distribution of the other and vice versa, such a coupling is modeled as a capacitive coupler $C_c$ between the two LC-circuits in the equivalent circuit diagram. Notice that the effective inductor $L_1$ associated with the GR SP resonance is mainly contributed by the kinetic inductance of the charge carriers [7,8].

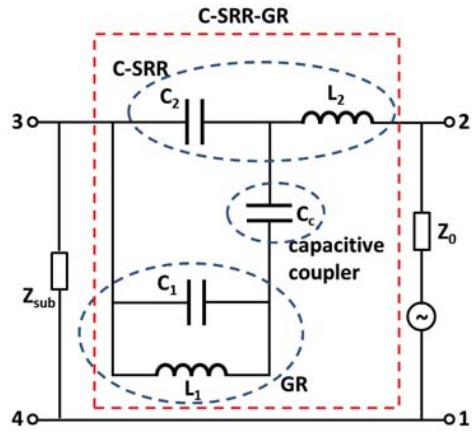

**Figure S5 | Equivalent circuit diagram of the C-SRR-GR hybrid metamaterial.**

Since the C-SRR is excited mainly by the incident electro-magnetic waves whereas the GR is driven mostly by the enhanced electric field in the capacitor gap of the C-SRR, the equivalent circuit for the entire system is constructed as shown in Fig. S5, with $Z_0$ and $Z_{sub}$ representing the impedance of the free space and the substrate, respectively. Therefore, the transmission peaks correspond to the zeros of the total impedance of the C-SRR-GR equivalent circuit measured between port 2 and port 3, which is straightforwardly derived to be

$$Z_{total}(\omega) = \frac{\omega^4 L_1 L_2 (C_1 C_2 + C_1 C_c + C_2 C_c) - \omega^2 (L_1 C_1 + L_2 C_2 + L_1 C_c + L_2 C_c) + 1}{-i\omega^3 L_1 (C_1 C_2 + C_1 C_c + C_2 C_c) + i\omega (C_2 + C_c)}.$$

(S15)

The two zeros of $Z_{total}(\omega)$ occur at

$$\omega_\pm^2 = \frac{L_1(C_1 + C_c) + L_2(C_2 + C_c)}{2} \pm \frac{\sqrt{(L_1(C_1 + C_c) - L_2(C_2 + C_c))^2 + 4L_1 L_2 C_c^2}}{2}.$$

(S16)

Let $L_1(C_1 + C_c) = \omega_1^2$, $L_2(C_2 + C_c) = \omega_2^2$, then Eq. (S16) becomes

$$\omega_{\pm}^2 = \frac{\omega_1^2 + \omega_2^2}{2} \pm \sqrt{\left(\frac{\omega_1^2 - \omega_2^2}{2}\right)^2 + \beta^2 \omega_1^2 \omega_2^2}, \quad \text{where } \beta = \frac{C_c}{\sqrt{(C_1 + C_c)(C_2 + C_c)}}.$$

(S17)

The above Eq. S17 reproduces Eq. 1 in the main text describing the anti-crossing behavior of two strongly coupled resonances, and the physical meaning of $\beta$ in this equivalent circuit model is the relative strength of the capacitive coupling.

## S5. Simulated carrier density dependent spectral response of the C-SRR-GR hybrid metamaterials operating at ~4 THz.

The simulated transmission spectra of the hybrid metamaterial device HM3 at various Fermi energies (hence carrier densities) and assuming two different carrier relaxation times (50 fs and 100 fs) are shown in Fig. S6, in comparison with that of the reference bare C-SRR array. The transmission spectra in Fig. S6a exhibit good agreement with the experimental observations in Fig. 5a. On the other hand, the transmission spectra in Fig. S6b clearly reveal the double-peak feature owing to the two hybridized modes in the strong coupling regime, similar to the experimentally observed spectral responses of the 10 THz devices in Fig. 3. From the spectra in Fig. S6b, a coupling strength of ~0.8 THz is extracted for the coupling between the C-SRR LC resonance and the GR localized SP resonance associated with this particular structure.

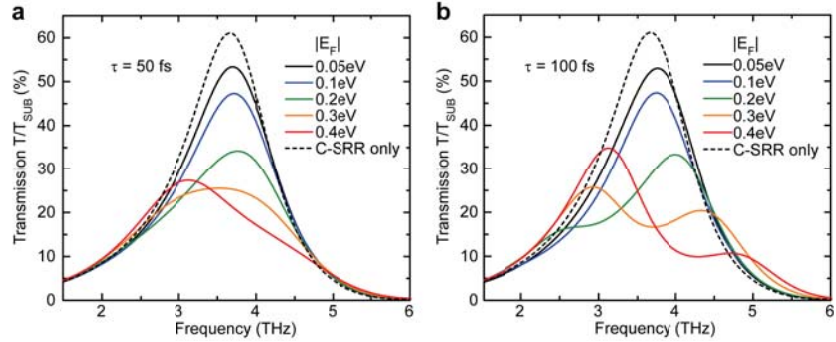

**Figure S6 | Simulated transmission spectra of tunable C-SRR-GR hybrid metamaterials operating at ~4 THz. a,b,** Simulated transmission spectra of device HM3 at various graphene Fermi energies ($|E_F|$) in comparison with that of the reference bare C-SRR array, assuming two different carrier relaxation time $\tau$, i.e. $\tau = 50$ fs (**a**) and $\tau = 100$ fs (**b**), respectively.

## S6. Additional analysis of the modulation speed measurement on the C-SRR-GR hybrid metamaterials.

The demonstrated modulation speed of the C-SRR-GR hybrid metamaterials is limited by the effective $RC$ time constant, contributed by both the device itself and the external driving circuit. The total capacitance of the device mainly stems from the capacitor formed between the Si substrate back-gate and the C-SRR-GR array, with the ~300 nm $SiO_2$ layer as the dielectric medium. Since the C-SRR-GR array covers most of the device surface area, the associated capacitance can be estimated using the parallel plate capacitor model, which yields ~115 pF for a 1 mm×1 mm device (HM3) and ~29 pF for a 0.5 mm×0.5 mm device (HM4). The effective total resistance $R$ is further extracted from the modulation speed measurement, and the extracted values are ~80 Ω for HM3 and ~180 Ω for HM4, respectively, excluding the 50 Ω input resistance of the driving voltage source. Thanks to the parallel electrical connection of all the C-SRR-GR unit cells, the resistance associated with the entire C-SRR-GR array is low (Fig. S7) and contributes to only a small fraction of the total resistance. Hence, the major contribution to the device total resistance is from the relatively low-doped Si substrate (resistivity ~10 Ωcm). Since electrical contacts for the current devices are made at the edges of the chip, the current flows in-plane in the Si substrate when charging and discharging the capacitor, resulting in large resistance. Such substrate resistance also depends on the area of the metamaterial field and its location on the chip. Since the measured device HM4 is located closer to the chip center than device HM3 and has smaller area, the substrate resistance seen by HM4 is higher. However, this substrate resistance contribution can be significantly decreased by more than one order of magnitude if electrical contacts are made on the back side of the thin Si substrate, which would facilitate the current to flow perpendicularly instead of in-plane. Such a back contact can be realized with a properly designed wire-grid structure, imposing minimal influence on the transmission of the THz radiation. Alternatively, one can also make use of such hybrid metamaterial structures to modulate the reflection of incident THz radiation, and in this case the Si substrate can be doped much higher.

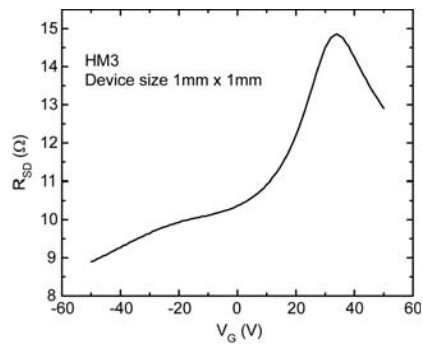

**Figure S7 | Measured source-drain resistance ($R_{SD}$) of the entire C-SRR-GR array (HM3) as a function of the back-gate voltage ($V_G$).**